\documentclass[reprint,
superscriptaddress,
 amsmath,amssymb,
 aps,
prx,
]{revtex4-2}

\usepackage{graphicx}
\usepackage{dcolumn}
\usepackage{bm}
\usepackage{hyperref}
\usepackage[mathlines]{lineno}
\usepackage{graphicx}
\usepackage{multirow}
\usepackage{array}
\usepackage[T1]{fontenc}
\usepackage{xcolor}

\begin{document}

\title{Role of tissue fluidization and topological defects in epithelial tubulogenesis}

\author{Richard D.J.G. Ho}
\affiliation{
Njord Centre, Physics Department, University of Oslo, Norway}%

\author{Stig Ove Bøe}
\affiliation{
Department of Microbiology, Oslo University Hospital, 0372, Oslo, Norway}%

\author{Dag Kristian Dysthe}
\affiliation{
 Njord Centre, Physics Department, University of Oslo, Norway}%

\author{Luiza Angheluta}
\affiliation{
 Njord Centre, Physics Department, University of Oslo, Norway}

\begin{abstract}
Cellular rearrangements, as primary sources of tissue fluidization, facilitate topological transitions during tissue morphogenesis. We study the role of intrinsic cell properties such as cell polarity and cell-cell adhesion in shaping epithelial tissues using a minimal model of interacting polarized cells. The presence of a vortex in the cell polarization poses the topological constraint that induces an inwards migration with the formation of a conical shape. Local rearrangements at the tip of the cone lead to the onset of tube formation. Switching between collective migration and structural rearrangements is key for balancing the contrasting tendencies, such as the tissue rigidity needed to preserve shape and the tissue fluidity allowing for topological transitions during tissue morphogenesis. 
\end{abstract}

\maketitle

\section{\label{sec:intro} Introduction}

Collective structural arrangements and migration of cells within epithelial tissues are important physical processes that underlie various phenomena such as embryonic or tissue development, wound healing, homeostasis and cancer metastasis. Understanding the complexity of cellular interactions and the resulting emergent collective behaviors presents formidable challenges both experimentally and theoretically. Epithelial tissues are a ubiquitous type of biological tissue found throughout the body of multicellular organisms. Epithelia are organized as mono- or multi-layered cell sheets that function as protective layers covering internal organs, line body cavities, and form the outer skin layer. 

On the structural level, epithelial monolayers can exhibit two distinct types of cell polarity: i) the apical-basal (AB) polarity, which provides the surface orientation, and ii) the planar cell polarity (PCP), orthogonal to the AB polarity axis that refers to the structural arrangement of cells within the tissue surface ~\cite{drubin1996origins,zallen2007planar}. 
Apical-basal polarity represents the main polarity axis, defining the ``top'' (apical surface) and ``bottom'' (basal surface) of the epithelial layer. Typically, the apical side faces the lumen or the outer surface of the monolayer, while the basal side is in contact with the basal membrane, which faces the underlying connective tissue \cite{apodaca2012role}. Planar cell polarity, on the other hand, is specified by a group of proteins specifically localized to the “front” and “back” (or “left” and “right”) of each cell, leading to structural alignment of cells within the monolayer~\cite{butler2017planar}. The functional role of PCP is evident in fully developed skin tissues, where PCP helps align hair follicles~\cite{montcouquiol2003identification,wang2006order}. In addition, PCP proteins may also play a role in dynamic processes during tissue development \cite{cetera2015round, stedden2019planar, classen2005hexagonal,gong2004planar,nishimura2012planar}. Epithelial monolayers can also undergo spontaneous polarization and collective migration in a manner that seem to be independent of PCP proteins. This process typically involves structural rearrangements of actin filaments, forming protrusions, called lamellopodia, at the leading edge of the cell. Such spontaneous transition is fundamental to the epithelial-to-mesenchymal transition (EMT), which plays a crucial role in wound healing, development, and tumor progression\cite{francou2020epithelial}.

Another characteristics of epithelial monolayers is their ability to maintain tight inter-cellular connections through cell-cell adhesion. This is essential for their barrier function and for their ability to control transport across the tissue. However, under specific conditions involving dynamic tissue remodeling, these epithelial sheets exhibit an ability to modulate cell-cell adhesion. By temporarily loosening cell-cell adhesion, they facilitate increased neighbor exchange and rearrangement, enabling necessary tissue movements. Examples of this kind of epithelial fluidization occur during wound closure~\cite{tetley2019tissue} and tissue morphogenesis and development~\cite{hannezo2022rigidity,petridou2021rigidity,pinheiro2024pulling}.

Epithelial monolayers have the ability to form intricate three-dimensional morphological structures such as folds, tubes, and branching networks~\cite{cetera2015round,stedden2019planar}. During morphogenesis, topological defects with full-integer charges emerge as key organizational centers for morphological events guiding topological transformation in evolving shapes. This is evidenced by experimental studies in the model organism Hydra, where long-lived $+1$ topological defects are formed through the nematic ordering of actin filaments and facilitate epithelial morphogenesis~\cite{maroudas2021topological}. In addition, a growing body of empirical evidence from various tissue systems suggests that similar topological defects also appear in the velocity patterns or structural alignment of cells within epithelial tissues during development and homeostasis. One prominent example is found in the small intestinal epithelium, characterized by the presence of crypts and villi. During homeostasis, epithelial cells within the crypt converge outwards and form a $+1$ defect in their structural arrangement. Similarly, at the villi tips, where cells migrate radially before shedding, effectively forming $+1$ defects by cells inwardly migrating towards the defect core \cite{gehart2019tales,perez2021mechanical,ritsma2014intestinal}. Furthermore, many epithelial tissues, such as those in the kidney, lung, and glands, exhibit branching and tubular morphologies. The initiation of these structures is thought to involve the formation of buds that extend orthogonal from the parent tissue, potentially creating a $+1$ defect within the velocity field at the budding site~\cite{eilken2010dynamics}. In addition, endothelial tissue, a specialized epithelium lining the vascular system, also exhibits similar branching morphology, indicating a potential  role of $+1$ defects in sprouting and blood vessel generation during angiogenesis~\cite{eilken2010dynamics,hendriks2020blood}.
Finally, the role of $+1$ defects in shaping epithelial morphogenesis is supported by studies showing  that epithelial monolayers cultured on flat surfaces spontaneously form vortices and swirls, signifying flow patterns that generate $+1$ defects in the velocity field \cite{doxzen2013guidance,heinrich2020size,li2014coherent,laang2022mechanical,segerer2015emergence}. 

Despite strong evidence suggesting a role for $+1$ defects in epithelial morphogenesis, several key questions remain unanswered, in particular what types of collective behaviors induce the formation of these defects and their feedback on morphological changes and tissue flows.

In this paper, we study the role of topological constraints and tissue dynamics on the tissue morphology. We approach this problem using a theoretical model of a tissue monolayer formed by polarized cells interacting with each other through cell-cell adhesion forces. We predict that during morphogenesis, the epithelial tissue evolves towards balancing competing tendencies, namely tissue rigidity needed for its functional integrity and sustaining shapes, and tissue fluidity enabled by structural rearrangements and needed for shape transformations. Within our model, the tissue rigidity is intrinsically rendered by the collective ordering of planar cell polarities, while in biological tissues there might also be environmental factors such as the cytoskeleton which may lead to the same effect. We find that a vortex in the tissue polarization induces a relaxational dynamics towards a conical shape. However, since the cone harbors a singular point, it is a source of structural rearrangements thus local fluidization which
triggers the onset of tube formation. The straight tube is sustained by the concentric pattern in the planar cell polarities with a fixed vortex. Fluctuations can break this radial symmetry and induce the formation a flap-like structure instead.


The rest of the paper is structured as follows. In Section \ref{sec:cellmodel}, we introduce a minimal model of polarized cells with adhesive interactions and discuss the finite size effects of the disk geometry. The cell migration patterns emerging around isolated full-integer defects in the orientation of planar cell polarities within a flat tissue are studied in Section \ref{sec:flat}. We show that asters and vortices induce transient outward and inward migration, respectively, whereas spirals sustain persistent spiraling migration. In Section~\ref{sec:curvedtissue}, we discuss the role of localized tissue fluidization in the transition from a cone to tubular structures. Concluding remarks and a summary are presented in Section~\ref{sec:conclusion}.      
\section{\label{sec:cellmodel} Minimal model of polarized epithelium}

\begin{figure}
    \centering
    \includegraphics[width=0.8\linewidth]{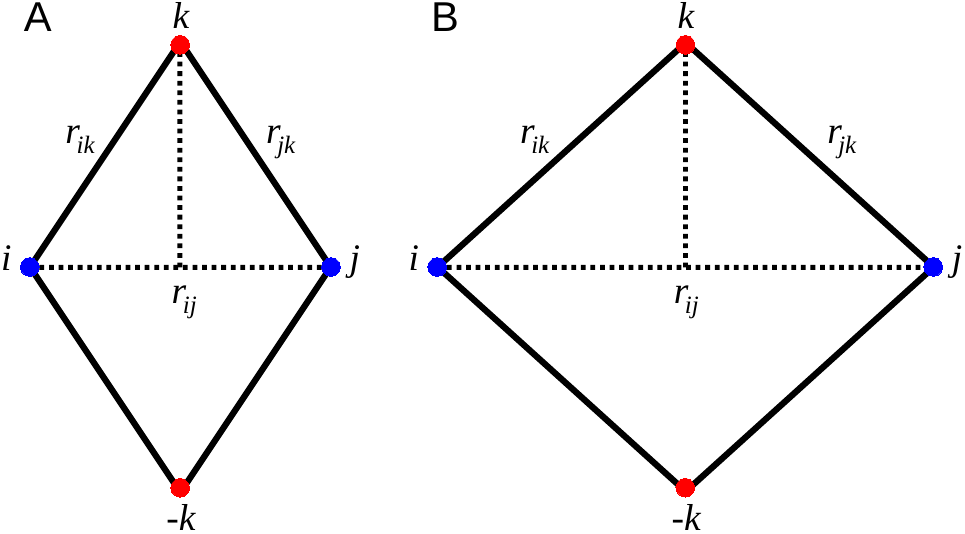}
    \caption{ A) Two neighboring cells, $i$ and $j$, since for all other third cells, $k$, $r_{ij} < r_{k-k}$, thus $r_{ij}^2 < r_{ik}^2 + r_{jk}^2$ and the angle $\angle IKJ < \pi / 2$.
    B) Two cells that are not neighbors, since there exists a third cell, $k$, where the angle $\angle IKJ > \pi / 2$ ($\lambda_0 = 2$). 
    }
    \label{fig:neighbour_calculation}
\end{figure}

To explore the interplay between tissue polarization and fluidization, we introduce a minimal model of polarized cells that interact with their neighbors through adhesion forces modulated by the cell polarities 
\cite{nissen2017four,nissen2018theoretical}. The Apical-Basal (AB) polarity $\mathbf p$ is a unit vector which is along the up-down axis of a cell and represents the cell elongation out of the tissue surface. Hence, $\mathbf p$ tends to point in the same direction as the outward normal of the tissue surface. Orthogonal to this is the planar cell polarity (PCP) represented by the unit vector $\mathbf q$ which lies within the tissue surface. The cells interact with their nearest neighbors (n.n) through a potential energy that contains two contributions: i) an isotropic interaction $V_{ij}^{(0)}$ that depends only on the pair separation distance, $r_{ij} = |\mathbf r_i-\mathbf r_j|$, and ii) an anisotropic part $V_{ij}^{(1)}$ that account for soft repulsion/attraction forces modulated by the local orientation of the cell polarities relative to each other and to their pair separation vector. Hence,  
$V_{ij} = V_{ij}^{(0)} + V_{ij}^{(1)}$, where~\cite{nissen2017four}
\begin{equation}
V_{ij}^{(0)} = \lambda_0 \left(e^{-r_{ij}} - e^{-r_{ij} /a_0}\right),
\end{equation}
with $a_0>1$, and minimized by an equilibrium pair distance $r_0 = \frac{a_0}{a_0-1}\ln a_0$. Adhesion forces mediated by polarities are anisotropic and described by the second contribution 
\begin{equation}
V^{(1)}_{ij} = e^{-r_{ij}} - (\lambda_1 S_1 + \lambda_2 S_2 + \lambda_3 S_3) e^{-r_{ij} / a_0},
\end{equation}
which is minimized by the same equilibrium distance when $\lambda_1+\lambda_2+\lambda_3=1$. The strength of the attraction is modulated by the orientation of polarities relative to each other and the pair separation vector,  
\begin{eqnarray}
S_1 &=& (\mathbf p_i \times \mathbf r_{ij}) \cdot (\mathbf p_j \times \mathbf r_{ij}) \\
S_2 &=& (\mathbf p_i \times \mathbf q_{i}) \cdot (\mathbf p_j \times \mathbf q_{j}) \\
S_3 &=& (\mathbf q_i \times \mathbf r_{ij}) \cdot (\mathbf q_j \times \mathbf r_{ij}),
\end{eqnarray}
such that orthogonal vectors are favoured for equilibrium configurations. We may notice that $S_1$ couples the surface normal with the cell positions, and maintains cells within the tissue surface whereby the AB polarities align along the normal to the tissue surface. On the other hand, the $S_3$ coupling the cell positions with the planar cell polarities may contribute to the normal migration of the cells. The $S_2$ term is important for maintaining orthogonality between the AB and PCP polarities, i.e. having two distinct polarities. The coefficient corresponding to $S_1$ needs to be greater than that of $S_3$ in order to identity the AB polarity with the surface orientation as discussed later. 

\begin{figure}
    \centering
    \includegraphics[width=0.85\linewidth]{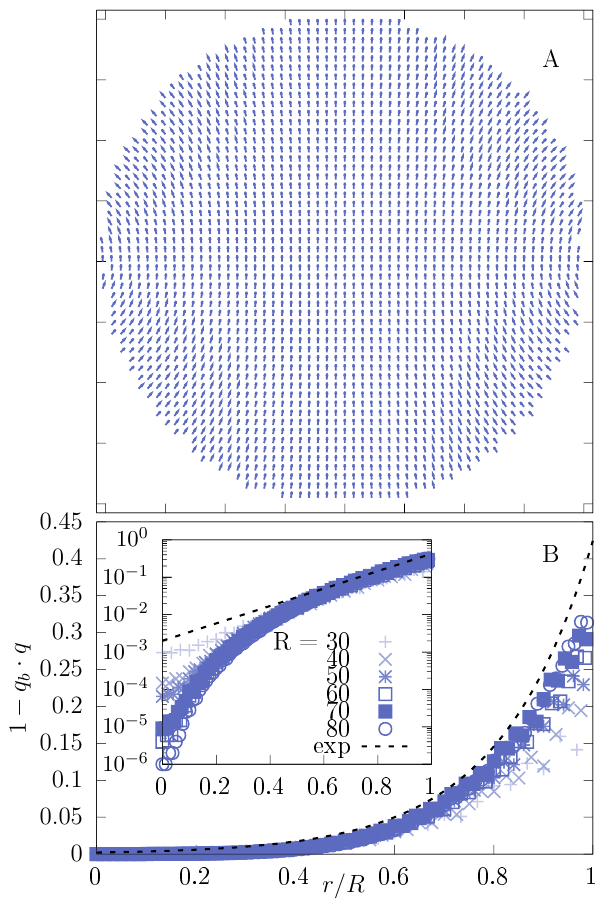}
    \caption{ A) Metastable orientation of $\mathbf{q}$. 
    B) Envelope of the deviation of $\mathbf{q}$ orientation from the uniform bulk for different system sizes. Semi-log inset shows the exponential relaxation with the characteristic lengthscale fitted to $\xi \approx 0.2 R$.  
    }
    \label{fig:skindepth}
\end{figure}

Cells migrate and reorient their polarities due to these interactions, following the 
overdamped dynamics 
\begin{eqnarray}
\dot{\mathbf{r}}_i &=& - \sum_{j =\textrm{n.n. of }i}\frac{\partial V_{ij}}{\partial \mathbf r_i} + \boldsymbol{\eta}^{(1)}_i \label{eq:r_n_dynamics}\\ 
\dot{\mathbf{p}}_i &=& - \sum_{j =\textrm{n.n. of }i}\frac{\partial V_{ij}}{\partial \mathbf p_i} + \boldsymbol{\eta}^{(2)}_i \label{eq:p_n_dynamics}\\ 
\dot{\mathbf{q}}_i &=& - \sum_{j =\textrm{n.n. of }i}\frac{\partial V_{ij}}{\partial \mathbf q_i} + \boldsymbol{\eta}^{(3)}_i, \label{eq:q_n_dynamics}
\end{eqnarray}
with small random fluctuations $\boldsymbol{\eta}^{(\alpha)}$ from the environment, modelled as uncorrelated Gaussian noise of zero mean and small standard deviation
\begin{equation}
    \langle {\eta}_{i}^{(\alpha)}(t) {\eta}_{j}^{(\beta)}(t')\rangle = 2\sigma \delta(t-t')\delta_{ij}\delta_{\alpha\beta}.
\end{equation}

The deterministic evolution preserves the unit norm of the polarities. However, due to noise, the norms can fluctuate. Therefore, we normalize the polarity vectors after each timestep.  

\begin{figure}
    \centering
    \includegraphics[width=0.85\linewidth]{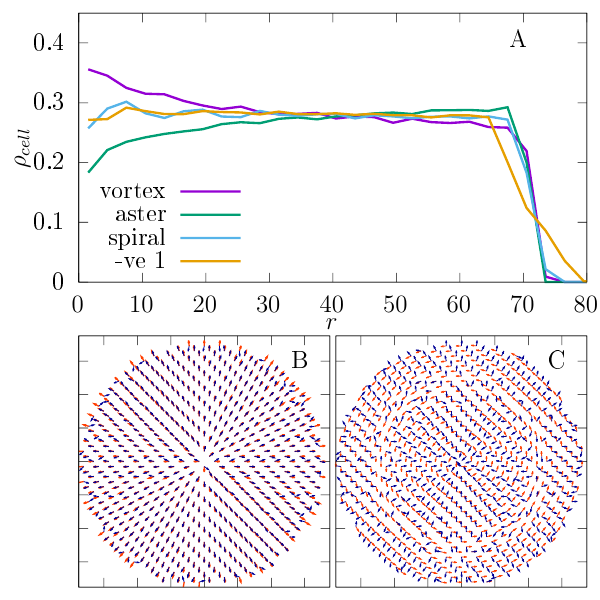}
    \caption{A) Radial dependence of the average cell density at a late time $t = 1000$ for different types of defects. 
    B) Instantaneous cell migration profile (blue) for an aster (B) and a vortex (C). Averages are taken over the transient time, $10$ initial configurations, and noise seeds. The corresponding polarity profiles are plotted in red. ($\lambda_0 = 2$)}\label{fig:flow_defect}
\end{figure}

\paragraph{Cell rearrangements:} 

We use an Euclidean metric to construct the network of cell neighbors. Cells $i$ and $j$ are considered neighbors if and only if they are closest in distance to their midpoint than any other third cell. This rule is achieved when the distances between cells $i$ and $j$ and any other cell $k$ satisfy the cosine rule $r_{ij}^2 < r_{ik}^2 + r_{jk}^2$, as illustrated in Fig.~(\ref{fig:neighbour_calculation}). This condition is similar to the Voronoi construction, but it is computationally more efficient and can be generalized to three-dimensions, and beyond single surfaces.

We distinguish two different dynamical regimes: i) cells remain in contact with their initial neighbors thus rendering tissue rigidity, and ii) cells can undergo neighbor exchanges to maintain the minimum pair distance, resulting in tissue fluidization. Numerically, it suffices to update the cell neighbor list at regular time intervals which may be longer than the discretization timestep. 

The evolution equations from (\ref{eq:r_n_dynamics})-(\ref{eq:q_n_dynamics}) with fixed cell-cell connectivity represent a relaxational dynamics towards minimising the total potential energy stored by surface bending (curvature) or orientational distortions in the alignment of planar cell polarities. In this regime, the collective cell migration tends to get arrested in metastable configurations due to frustrations in the cell connectivity network. This is, however, circumvented by relative migration corresponding to local rearrangements to preserve the minimum neighbor distance.

In this computational study, we set 
the parameters associated with the anisotropic interactions to fixed values given by $\lambda_1= 5/11$, $\lambda_2 = 4/11$ and $\lambda_3= 2/11$ such that they sum up to one. The relative values of $\lambda_1$ and $\lambda_3$ determine which of the polarities point out of the surface and which remain in the surface. For $\lambda_1>\lambda_2$, we ensure that the $\mathbf{p}$'s represent the AB polarities pointing out of the surface. We have included the specific value of $\lambda_0$ for the isotropic interactions in each figure caption. 
The value of $\lambda_0$ determines the strength of the isotropic interactions in comparison to the polar forces. Since we keep $\lambda_1+\lambda_2+\lambda_3=1$, a value $\lambda_0 \ll 1$ means that isotropic adhesion is negligible, whilst $\lambda_0 \gg 1$ leads to clumping. The isotropic interactions are important for maintaining the tissue as a simply connected surface under outwards migration, like the one induced by an aster (no holes).

We simulate the dynamical equations using as initial condition a disk geometry with hexagonal packing of the cells. The cell positions and polarities have free boundary conditions. For a flat tissue, we fixed the AB polarities along the z-axis, i.e. $\mathbf p = (0,0,1)$. This suffices to maintain the evolution of the cells within the $(x,y)$ plane. The different configurations of cell polarities are discussed separately in the next sections. 


\paragraph{Finite-size effects:} We use free boundary conditions for the dynamical variables. To quantify the boundary effect on the PCP polarities, we consider a metastable configuration where the $\mathbf{q}$ polarities uniformly point in the same direction for the bulk cells and rotate in different directions for the cells within a boundary layer to ensure that there is a net $+2\pi$ rotation imposed by disk geometry through its Euler characteristic $\chi=1$. This is also illustrated in Fig.~(\ref{fig:skindepth} a)). We measure the local deviation in the orientation of the $\mathbf{q}$ polarities from their uniform (bulk) orientation $\mathbf{q}_b$, i.e. $1-\mathbf q_b\cdot \mathbf q_n$ as a function of the radial distance from the center of the disk domain. In Fig.~(\ref{fig:skindepth} B), the envelope of this deviation is plotted as function of the radial distance $r/R$ (relative to the radius $R$). The deviation decreases exponentially $e^{-(R-r)/\xi}$ with the distance from the boundary and its corresponding characteristic length $\xi\approx 0.2 R$ is independent of model parameters. The hexagonal packing of cells introduces a ragged edge of the disc which also influences the perpendicular orientation of the polarity at the disc edge. However, this effect is much more localized and negligible compared to the one introduced by the free boundary conditions. 
\begin{figure}
    \centering
    \includegraphics[width=.85\linewidth]{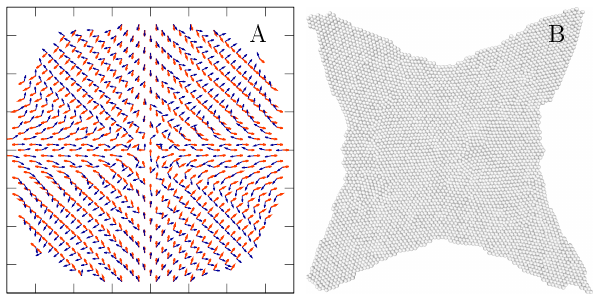}
    \caption{
    A) Transient cell migration induced by $-1$ defect deforms the disc into an "X" shape. Averages are taken over the transient time, initial conditions and noise seeds. 
    The structure of the polarities is shown in red.  
    B) Snapshot of the tissue configuration at $t = 5000$ 
    induced by the $-1$ defect. ($\lambda_0 = 2$) 
    }
    \label{fig:persistent_flow_defect}
\end{figure}
\section{\label{sec:flat} Migration patterns in a flat tissue}

Through cell-cell adhesion forces described by the Eq. (\ref{eq:r_n_dynamics})-(\ref{eq:q_n_dynamics}), the PCP polarities $\mathbf q_i$ tend to align along a preferred orientation to form polar order. Since the disk geometry has the Euler characteristic $\chi=1$, there will be a net $+ 2\pi$ rotation of the $\mathbf q_i$. This can be achieved as an edge effect (see Fig.~\ref{fig:skindepth}) or through a structural arrangement induced by the presence of a topological defect of $+1$ charge at the center of the disk.  Topological defects locally melt the orientational order and induce long-range deformations which feed into both collective and relative migration. 

We first consider a flat tissue with fixed AB polarities along the normal to the tissue plane, i.e. $\mathbf{p}_n = (0,0,1)$, and allow cells to migrate following Eqs.~(\ref{eq:r_n_dynamics}) with or without neighbor exchanges.

As proof-of-concept and to bridge with hydrodynamic models, we consider the setup of a single defect embedded in a uniform polarity orientation. The profile of the PCP polarities $\mathbf q_i$ induced by the defect is fixed instantaneously to the imposed orientational structure of an isolated defect. Alternatively, one may include an energy contribution due to deviations from the $\mathbf p_i$ polarities from the imposed profile, resulting in a linear restoring force to the desired polarity configuration. 
\begin{figure}
    \centering
    \includegraphics[width=0.85\linewidth]{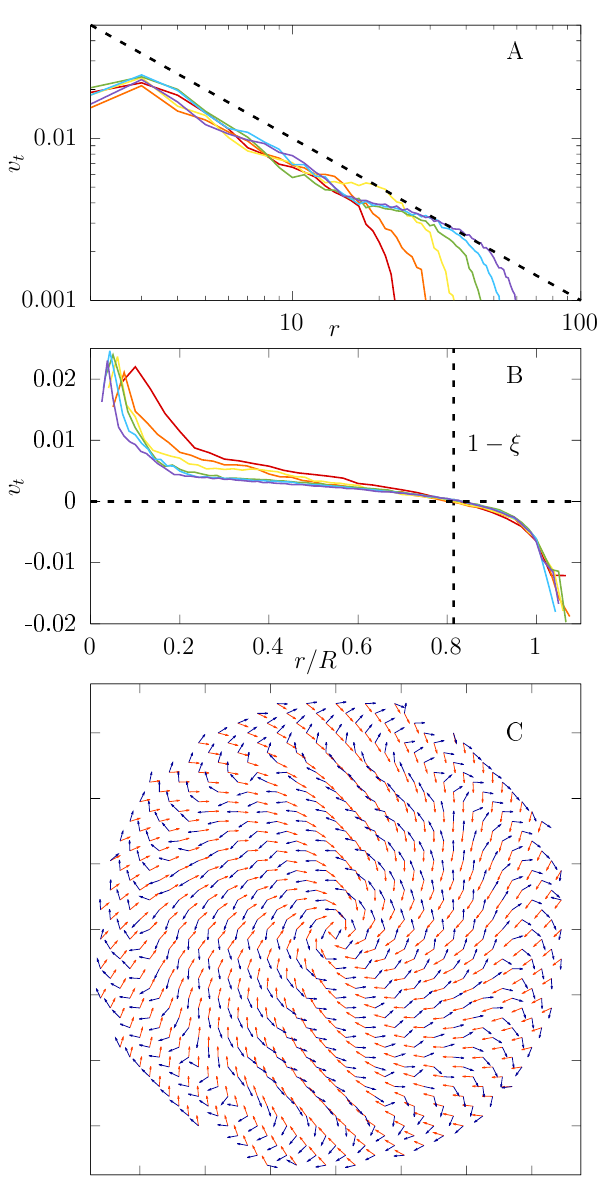}
    \caption{ A-B) Average azimuthal flow velocity as a function of the radial distance from a spiral ($m=1$, $\theta_0 = \pi/4$) for different disk radius, $R$. The dashed line shows the $1/r$ scaling. The finite size effect is determined by the thickness of the boundary layer $\xi$.
    C) Persistent flow for a spiral defect. Motion of the particles in blue, embedded defect in orange ($\lambda_0 = 2$).
    }
    \label{fig:vortical_velocity}
\end{figure}

The isolated defect inserted at the center of the disk imposes a polar orientational field $\theta$ given by \cite{ronning2023spontaneous}
\begin{equation}
\theta = \theta_0 + m \arctan{(y / x)} \ , \
\end{equation}
where $m=\pm 1$ is the topological charge picked up by an arbitrary contour integral $\oint\limits_C d\theta= 2\pi m$ enclosing the defect. The constant phase $\theta_0$ gives the baseline of the defect. For a defect with $m=-1$, the baseline phase can be set to $\theta_0=0$ by a reorientation of the defect. However, this baseline phase $\theta_0$ is an intrinsic phase for the $m=+1$ defect, and its value distinguishes three types of defects: $\theta_0 = 0$ corresponds to an aster, $\theta_0 = \pi/2$ for a vortex and $\theta_0 \in (0, \pi/2)$ gives a spiral defect~\cite{ronning2023spontaneous}. Due to their migrations, cells acquire different orientations of their PCP polarity depending on their distance from the defect. Thus, while the polar orientation $\theta$ is quenched, the actual $\mathbf{q}_i$ polarities are obtained by evaluating $\theta_i = \theta(\mathbf r_i)$ at their current cell position $\mathbf r_i$, i.e. $\mathbf{q}_i = (\cos(\theta_i),\sin(\theta_i),0)$. 

\subsection{Transient migration}
With or without neighbor exchanges, both an aster ($\theta_0=0$) and a vortex ($\theta_0=\pi/2$) give rise to a transient radial migration corresponding to an outward motion for an aster (regardless on whether the aster arrangement of cell polarities is inward or outward) and an inward motion for a vortex. This divergent/convergent migration leads to depletion/accumulation of cell density around an aster or a vortex, respectively, as shown the Fig. (\ref{fig:flow_defect}, A). We also compute the typical patterns of cell migration by averaging over different initial conditions, time and noise seeds. These average profiles are illustrated for an aster (B panel) and a vortex (C panel). The divergent migration from an aster is balanced out by the isotropic attraction forces (tuned by $\lambda_0$ parameter) preventing ruptures, such that the cells tend to stagnate into simply connected configurations. 

Notice that for an aster, the outward migration pattern aligns with the divergent polarity profile, while for the vortex there is a reversal of motion from inward in the bulk to outward on the rim of the disk. This reversal effect is characteristic to $+1$ defects as predicted from a hydrodynamic model in Ref.~\cite{ronning2023spontaneous}. Interestingly, we find that it is also present for a spiral defect as discussed later. 

\begin{figure}
    \centering
    \includegraphics[width=0.85\linewidth]{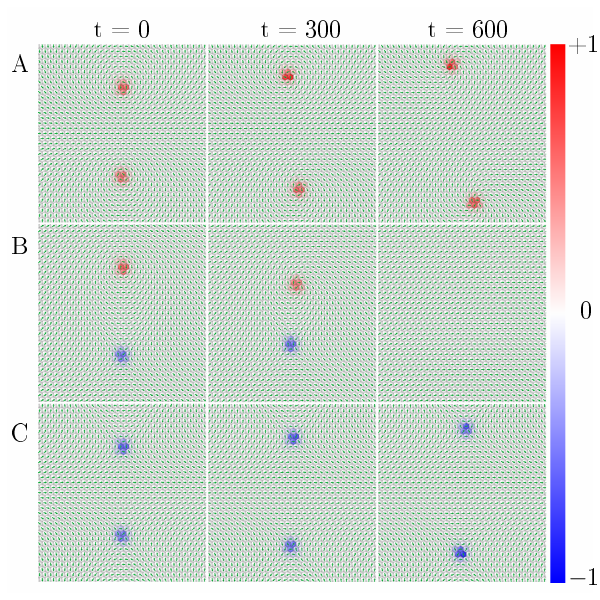}
    \caption{Time snapshots of the evolving $\mathbf{q}_i$ polarities for two topological defects: A) Two initial vortices turning into spirals while repelling each other, B) One vortex turning into a spiral as it approaches the $m=-1$ defect, C) Two $m=-1$ defects repelling each other. Background field shows the defect density charge $\rho$ field ($\lambda_0 = 1.5$). 
    }
    \label{fig:two_defects}
\end{figure}

For the $-1$ defect, the cell polarity alignment has a  4-fold saddle structure. This induces an 8-fold saddle structure in the migration pattern as shown in Fig.~(\ref{fig:persistent_flow_defect}, B) and consistent with the theoretical prediction from a hydrodynamic model of polar active matter~\cite{ronning2023spontaneous}. The saddle point introduces large frustrations in the cell-cell connectivity which are removed by local neighbor exchanges. This spontaneous rearrangements occur along the principal axes oriented at $\pi/4$ degrees with respect to the saddle point axes and leads to the transient fluidization with the formation of an "X" shaped tissue.

\subsection{Persistent migration}
The only configuration with persistent cell migration corresponds to a spiral $\theta_0\in(0,\pi/2)$. Under neighbor exchanges, cells sustain a chiral flow as shown in Fig.~(\ref{fig:persistent_flow_defect} A). The flow chirality is determined by that of the spiral in the center of the disk and changes sign near the boundary of the disk. This is similar to the reversal of transient divergent migration induced by a vortex and the flow reversal predicted in Ref.~\cite{ronning2023spontaneous}. To further bridge to the hydrodynamic models, we compute the azimuthal speed for this vortical flow as function of the radial distance to the spiral defect. In Fig.~(\ref{fig:vortical_velocity}), we see that it  exhibits the $1/r$ scaling at intermediate scales away from the defect core (discrete nature) and the finite boundary. Notice that the azimuthal speed reverses its sign near the boundary layer, i.e. at $1-\xi$, which implies that $\xi$ is the relevant lengthscale for this flow reversal. 

In Fig.~(\ref{fig:two_defects}), we illustrate the typical orientational flow fields (normalized vector field) induced by one pair of defects (two vortices, one vortex and one $-1$ defect and two $-1$ defects). The cell migration pattern changes non-trivially in the presence of multiple defects due to the long-range interactions between defects.  

\begin{figure}
    \centering
    \includegraphics[width=0.85\linewidth]{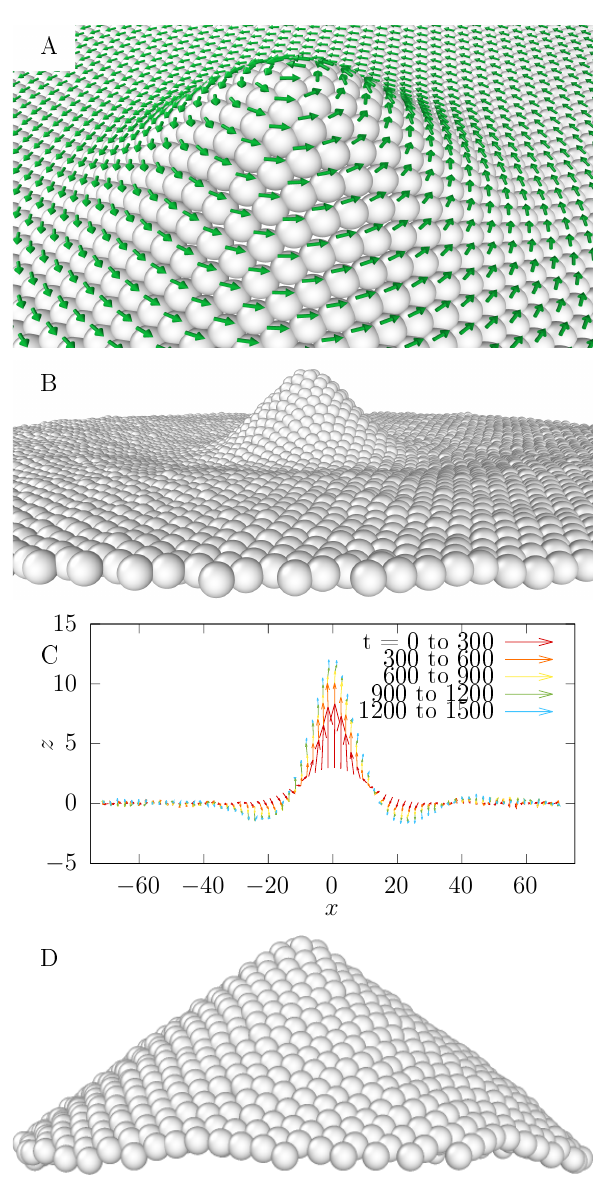}
    \caption{Morphological response to an imprinted vortex in the absence of neighbor exchanges.
    A) Initial Gaussian bump perturbation (with height exaggerated for visualization purposes).
    B) Formation of a conical shape.
    C) Profile of the instantaneous cell velocity along the $(x,z)$ cross-section for different times.
    D) Relaxation to the asymptotic conical shape ($\lambda_0 = 2$). 
    }
    \label{fig:gaussian_bump}
\end{figure}
\subsection{Defect density field}

To further characterize the polar order within the tissue and its topological defects, we coarse-grain the $\mathbf q_n$ polarities to obtain a smooth polarization field 
\begin{equation}
  \mathbf \Psi (\mathbf r') = \frac{1}{N}\sum\limits_{i=1}^N  \mathcal P_i\cdot \mathbf q_i \delta(\mathbf r'-\mathbf r'_i)
\end{equation}
corresponding to an $O(2)$ vector order parameter of the tissue polarization. $\mathbf r' = \mathcal P_i\cdot \mathbf r$ is the in-plane coordinate and $\mathcal P_i = (\mathcal{I}-\mathbf p_i\mathbf p_i)$ is the in-plane projection operator for the $i$-th cell. Whilst the $\mathbf q_i$ are unit vectors, the corresponding polarization field has varying magnitude (i.e. the order parameter space is the unit disk). The topological defects are associated with regions where the polarization vanishes in magnitude and becomes multi-valued in phase. Thus, defects can be tracked as zeros of the order parameter using the Mazenko-Halperin method~\cite{skogvoll2023defects}. The defect density $\rho$ follows as
\begin{equation}
\rho = \frac{1}{\pi \Psi_0^2}[(\partial_x \Psi_x)(\partial_y \Psi_y) - (\partial_x \Psi_y)(\partial_y \Psi_x)], \ 
\end{equation}
where $\Psi_0$ is the constant magnitude of the uniform polarization. Topological defects are initially embedded in the phase of the polarization field as singularities 
\begin{equation}
\theta = \theta_0 + \sum_i m_i \arctan \bigg( \frac{y-y_{0,i}}{x-x_{0,i}} \bigg) \ ,
\end{equation}
with topological charges $m_i = \pm 1$. For two defects, we use the initial position coordinates $(x_{0,i},y_{0,i}) = (0,\pm 15)$ and evolve the cell positions and the $\mathbf q_i$ polarities according to Eqs.~(\ref{eq:r_n_dynamics}) and (\ref{eq:q_n_dynamics}), where cells are allowed to exchange neighbors in order preserve the minimum neighbor distance. 

As shown in Fig. (\ref{fig:two_defects}), the defect density field $\rho$ tracks very well the topological  charges of the moving defects. The defect core size is about two cell units, thus small enough compared to the system size to observe hydrodynamic effects and long-range interactions between defects. We see that defects of same sign repels while those of opposite signs attract and eventually annihilate, as predicted in hydrodynamic models.

\begin{figure}
    \centering
    \includegraphics[width=0.85\linewidth]{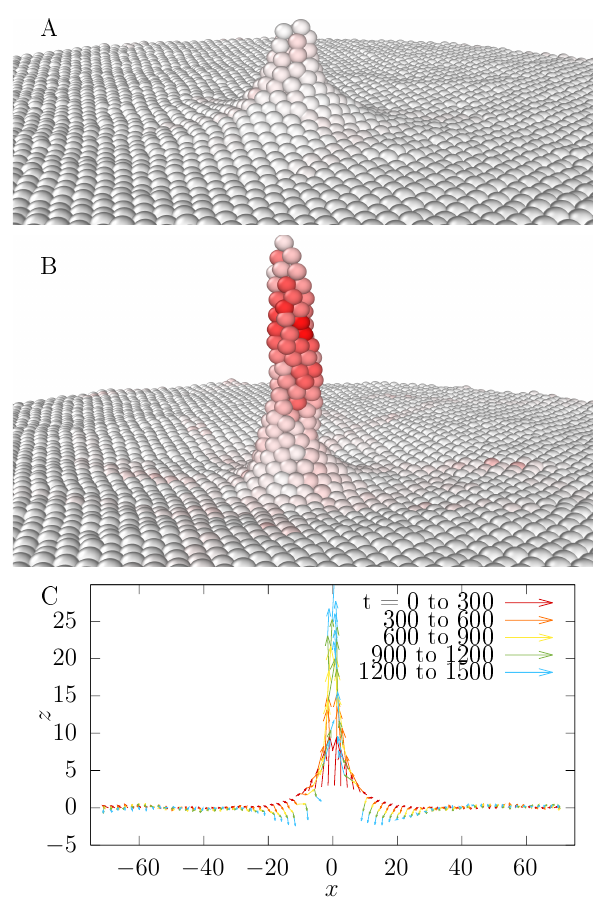}
    \caption{Morphological response induced by a fixed vortex in the presence of cell rearrangements. 
    A) Tubular growth at t=500, B) at t = 1500. C) Profile of the instantaneous cell velocity along the $(x,z)$ cross-section for different times. 
    The colormap shows the density of neighbor exchanges per cell indicating localized fluidization around the evolving tube ($\lambda_0 = 2$).}
    \label{fig:morphs_imprint}
\end{figure}

\section{\label{sec:curvedtissue} Cone versus Tube}

We now relax the zero curvature constraint by allowing the AB polarities to evolve according to Eq.~(\ref{eq:p_n_dynamics}), which will induce cell migration normal to the surface. We study the morphological response of the tissue due to presence of a vortex in the $\mathbf q_i$ polarities and see how cell neighbor exchanges induce the formation of a tubular structure. 

To induce a morphological change, we perturb the disk with a Gaussian bump of small height $h= 3$ cell units as illustrated in Fig.~(\ref{fig:gaussian_bump}, A). Using the radial symmetry of the vortex, we can extend the method of embedding a vortex in the configuration of the $\mathbf q_i$ polarities from Sec.~(\ref{sec:flat}) by fixing the orientation of the $\mathbf q_i$ along concentric rings centered at the vortex position. The cell migration and the evolution of tissue curvature are governed by the dynamics of $\mathbf r_i$'s and $\mathbf p_i$'s. At each timestep, we evaluate the orientation of the $\mathbf q_i$ polarities along concentric rings at a distance $r$ from the vortex using the static orientational profile of $\theta$. 
Embedding other types of full-integer defects for curved tissues is more challenging and remains to be studied separately. 

\begin{figure}
    \centering
    \includegraphics[width=0.85\linewidth]{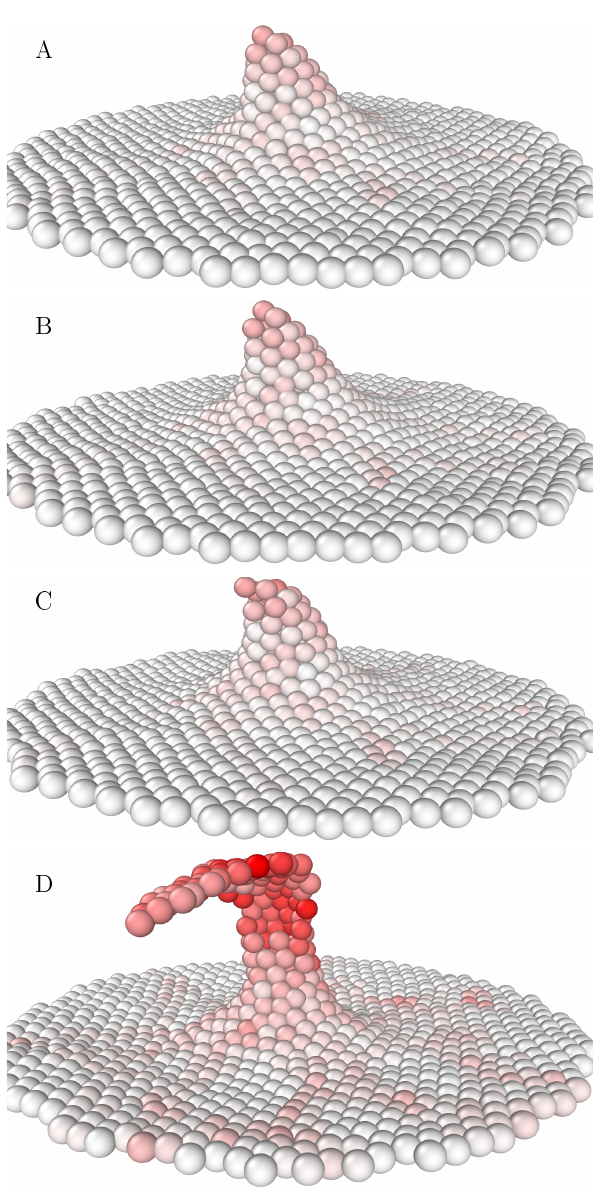}
    \caption{Time snapshots during the formation of a flap under vortex dynamics and cell rearrangements: A) $t=575$, B) $t=625$, C) $t=675$, D) $t= 1500$. 
    The colormap shows the density of neighbor exchanges per cell and indicates that structural reconfiguration is localized around the evolving flap ($\lambda_0 = 2$).}
    \label{fig:FlapEvolution}
\end{figure}

The vortex is a source of inward migration and accumulation of cells towards the Gaussian bump and this leads to a normal migration. From energetic considerations, the vortex localises the energy associated with large orientational distortions. In the absence of neighbor exchanges, the collective migration is towards minimizing this total energy and thus attaining an equilibrium shape. Subsequently, the normal migration is seeded at the vortex center by the "leading" cells that pull the rest of the tissue into a conical shape. The relaxation towards the cone is shown in Fig.~(\ref{fig:gaussian_bump} A-D). The overdamped dynamics slows down as the conical shape is approached asymptotically. This equilibrium shape is attained also when we allow the PCP polarities to evolve unconstrained.
\begin{figure}[t]
    \centering
    \includegraphics[width=0.85\linewidth]{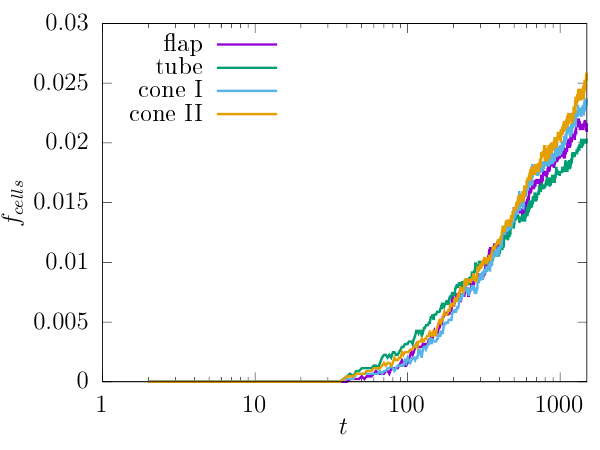}
    \caption{Onset of out-of-plane growth. 
    The fraction of cells that are more than two cell widths above the flat surface as a function of time in semilog-scale: flap, tube, cone I (rigid, moving defect) and cone II (rigid, fixed defect) ($\lambda_0 = 2$). }
    \label{fig:shape_growth}
\end{figure}
\begin{figure}[t]
    \centering
    \includegraphics[width=0.85\linewidth]{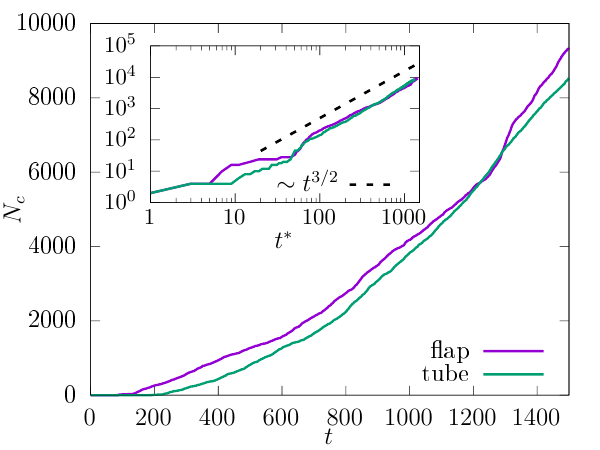}
    \caption{The number of neighbor exchanges, $N_c$, as function of time for tube and the flap shapes. Inset the dependence in the log-log scale after the initial elastic response ($\lambda_0 = 2$).}
    \label{fig:N_c_statistics}
\end{figure}
By allowing the local neighbor rule to change the cell neighbor connectivity, we enable localized fluidization, and this fundamentally alters the normal migration path. For the vortex configuration, where the inwards migrations within the tissue leads to normal migration, the local fluidization induces a topological transition from a conical shape to a tubular structure as shown in Fig.~(\ref{fig:morphs_imprint}).

We introduce a scalar field defined by the cumulative number of neighbor exchanges per cell to better quantify the fluidization at the onset of tube formation and show that it is highly localized in space. This is shown in the colormap of Fig.~(\ref{fig:morphs_imprint}). The density of rearrangements is zero in the beginning as the tissue responds by collective relaxation towards the conical shape. Eventually, the curvature at the tip of the cone is sufficiently high to trigger a local cell neighbor exchange which blunts the tip by opening a small hole and marks the onset of tube formation as shown in Fig.~(\ref{fig:morphs_imprint}). The continuous cell rearrangements at the top and on the side of the tube maintains the normal growth. Most of the cells away from the tube remain fixated by their neighbors and migrate collectively. However, the fluidization initiated by the normal growth permeates through the in-tissue migration on localized streaks or shear zones. 

Interestingly, the unconstrained evolution of the $\mathbf q_i$'s polarities  renders a morphological transition from a tube to flap structure as shown in Fig.~(\ref{fig:FlapEvolution}). At the base of the bump and in the far-field, the concentric ordering of the polarities induced by a vortex is still maintained, but the vortex center may move its position due to noise. These fluctuations are enough to spontaneously break the rotationally-symmetric shape by inducing a bend. This leads to a difference in curvature between the inside edge of the bending tube (higher curvature) and the outside edge of the tube (lower curvature).
As a consequence, the side of tube with higher curvature stores more energy and stagnates in the growth, whilst the side of the tube with lower curvature continues to grow rendering a flap-like shape. Interestingly, the ordering of the polarities on the flap retains the vortex-like structure. This indicates that the tissue would rather change its shape than alter its topological ordering, pointing to the robustness of the topological constraint. The density of structural rearrangements remains localized on the growing flap and along narrow steaks in the far-field akin to shear zones in granular matter (see Fig. \ref{fig:FlapEvolution}).

This topological robustness is also reflected in the global properties of the growing shape and the total number of neighbor exchanges as shown in Figs.~(\ref{fig:shape_growth}) and (\ref{fig:N_c_statistics}). At the onset of normal growth, we see that the density of cells inside the growing shape tends to increase logarithmically $\sim \ln t$  for different classes of shapes with or without neighbor exchanges. We attribute this logarithmic growth to the presence of the vortex guiding the growth through the inward migration pattern, but this requires further theoretical study. On the other hand, the total number of the neighbor exchanges tends to increase algebraically with time as $t^{3/2}$ in the asymptotic limit. This exponent is the same for both the tube and flap and indicates that it is determined by the topological constraint imposed by the vortex. It also implies that the rate of rearrangements, representing the flux of cells into the fluidized (shear) zones, scales as $\sim t^{1/2}$, from the time derivative of the density of neighbor exchanges.

\begin{figure}[t!]
    \centering
    \includegraphics[width=0.85\linewidth]{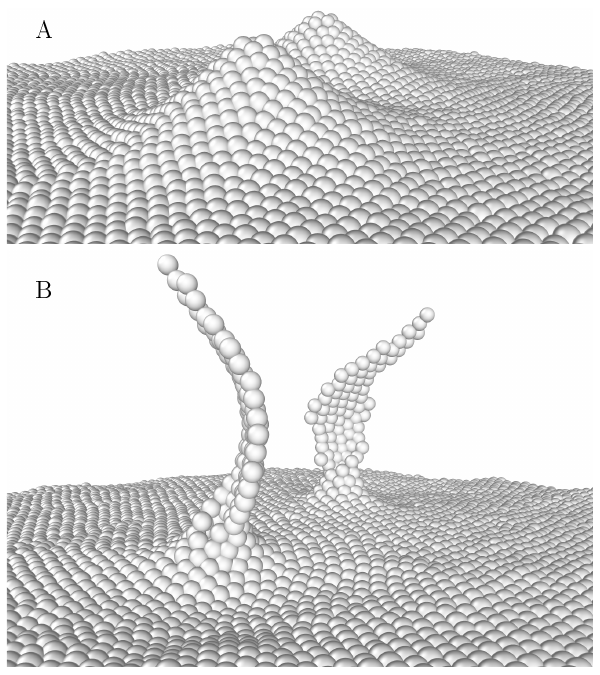}
    \caption{Morphological changes of two initial Gaussian bumps due to two interacting vortices. A) Conical shapes in the solid-limit, B) flap-like shapes during tissue fluidization ($\lambda_0 = 2$). 
    }
    \label{fig:two_bumps}
\end{figure}

To further emphasise the topological robustness, we consider the morphological change induced by two vortices. For an intact tissue rigidity, we find that the two vortices lead to collective normal migration with the formation of two conical shapes as shown in Fig.~(\ref{fig:two_bumps} A). By contrast, two flaps are formed under tissue fluidization and when vortices are allowed to move away from each other due to their repelling interaction as shown in Fig.~(\ref{fig:two_bumps} B). We notice that as the vortices turn into spirals as in Fig.~(\ref{fig:two_defects}), the flaps also develop the same chirality as that of the corresponding spiral at the base.

\section{\label{sec:conclusion} Discussion and conclusions}

In summary, we have studied the interplay of topological defects and tissue fluidization on the morphological transformations of a tissue monolayer using a minimal discrete
model of interacting polarized cells. Topological defects in the ordering of planar cell polarities pose robust topological constraints on tissue shape. However, we find that the collective migration, which is preferred by epithelial cells, gets fragmented by the localized tissue fluidization during shape dynamics. In particular, we have shown that under the topological constraint imposed by a vortex, the inwards migration within the
tissue leads to a normal migration in the region of high curvature. We find that the onset of tube formation from a conical shape is attributed with the onset of tissue fluidization occurring at the tip of the cone. Local
rearrangements may trigger each other similar to shear transformation zones in amorphous materials leading to the formation of shear zones where the tissue yields and flows. 

The observed flap structures, extending from the tubular protrusion of the epithelial surface, suggest a mechanism by which cells at the tip can be displaced from the contiguous monolayer plane. This results in a cohort of cells at the protrusion with both sides exposed to the extracellular environment, potentially facilitating enhanced interaction with their surroundings. A similar phenomenon occurs during angiogenesis, the formation of new blood vessels. Specialized tip cells, known
as endothelial sprouts, breach the contiguous endothelial
monolayer through a process called sprouting. Subsequent cell division and pulling by neighboring cells lead to the formation of a collective of specialized endothelial cells at the sprout tip. These tip cells exhibit invasive properties, enabling the growth of the new blood vessel~\cite{eilken2010dynamics,hendriks2020blood,geudens2011coordinating}.

While several studies have shown the ability of developing tissues to regulate their cellular fluidity during morphogenesis, the role of fluidization in epithelial tubulogenesis remains largely unexplored. Interestingly, a recent work highlights the potential role of regulating tissue fluidity in the spiraling alignment of endothelial cells within the tubular structure of blood vessels~\cite{zhang2024helical}. Thus, our computational model, which demonstrates an increased propensity for local fluidity during epithelial tube formation, suggests that the regulation of tissue fluidity might be a general and critical process during epithelial tube morphogenesis. 

Currently, the model does not integrate the influence
of the micro-environment, which comprises essential extrinsic factors like the extracellular matrix (ECM), signaling molecules, and mechanical cues. Additionally, the
model does not consider cellular processes such as cell
division, extrusion, and apoptosis, which could be vital
for remodeling and homeostasis of the epithelial \emph{in vivo}. 
Yet, we have shown that this minimal model offers a valuable advantage by capturing the essential, intrinsic mechanisms governing epithelial cell sheets, such as polarity and inter-cellular adhesions. This suggests that intrinsic cell properties, combined with topological constraints (which may also be posed by the micro-environment) which trigger the tissue response to converge towards a $+1$ defect core, might be sufficient for building epithelial tubes. Thus, the present study highlights the importance of fundamental cell-intrinsic principles in shaping complex tissues. Further investigations could explore how these intrinsic properties of epithelial monolayers cooperate with extrinsic factors, such as mechanical cues and biochemical signaling, to facilitate the formation of complex organ structures. These provide many future directions of research for this model, with the present work serving as a basis.

\begin{acknowledgments}
This project was funded by the University of Oslo, UiO:Life Science through the convergence environment ITOM.

\end{acknowledgments}


\bibliography{refs_monolayer}

\end{document}